\documentclass{article}

\usepackage[final]{neurips_2019}

\usepackage[utf8]{inputenc} 
\usepackage[T1]{fontenc}    
\usepackage{hyperref}       
\usepackage{url}            
\usepackage{booktabs}       
\usepackage{amsfonts}       
\usepackage{nicefrac}       
\usepackage{microtype}      
\usepackage{graphicx}
\usepackage{subfigure}
\usepackage{algorithm}
\usepackage{algorithmic}
\title{Assessing Viewer’s Mental Health by Detecting
Depression in YouTube Videos}

\author{%
  Shanya Sharma\\
  SAP Labs\\
  \texttt{sharmashanya1297@gmail.com} 
   \And
   Manan Dey \\
   SAP Labs \\
   \texttt{manandey01@gmail.com} \\
}

\begin{document}

\maketitle
\begin{abstract}
Depression is one of the most prevalent mental health issues around the world, proving to be one of the leading causes of suicide and placing large economic burdens on families and society. In this paper, we develop and test the efficacy of machine learning techniques applied to the content of YouTube videos captured through their transcripts and determine if the videos are depressive or have a depressing trigger. Our model can detect depressive videos with an accuracy of 83\%. We also introduce a real-life evaluation technique to validate our classification based on the comments posted on a video by calculating the CES-D scores of the comments. This work conforms greatly with UN Sustainable Goal of ensuring Good Health and Well Being with major conformity with section UN SDG 3.4.
\end{abstract}

\section{Introduction}
\label{submission}

Depression is affecting as high as 300 million people around the world, according to the World Health Organization. However, the WHO reports that a clear majority of depressed individuals never seek out treatment, mostly because of being unaware of what is going on with them [11] .

We propose a solution to aid in early diagnosis of depression by giving a zoomed-in perspective of a  patient’s state of mind and experiences in order to improve diagnostic accuracy and efficiency. Our solution is to develop a model with the ability to track an individual’s YouTube watching activity and predict his/her state of mind by analyzing the patterns in the kind of videos watched by him/her over a period of time. We realize that a false positive result from the same can lead to anxiety and other psychological impacts, therefore we also propose a procedure to validate the results produced by the model using CES-D scale [15] used for screening of depression by primary health care and in research. A basic flow describing our solution is given in Figure 1.

Unlike other work that aims at detecting depression from an individual’s activity on social media platforms  [9], [12], [3], [1], [17], [10], [4] examining clinical interview videos [7], [2], [14] where posts/interviews by an individual were self-descriptive, we aim at collecting data passively from user’s end to get an objective view of his/her mental state. 

At a high level, our pipeline is composed of 3 different layers: i) Secured data acquisition, ii) Feature extraction from individual videos and their content analysis iii) Analysis of video viewing patterns over time. Examples of the types of patterns we aim to find are:

\begin{itemize}
\item{A gradual decline from happy/uplifting videos to depressing/sad videos.}
\item{Videos of same emotional content, since they tend to do repetitive tasks for a long period of time (rumination).}
\item{High frequency of depressive videos in the watching history}
\end{itemize}

\begin{figure*}[h]
\begin{center}
\includegraphics[width = 4.8in, height = 1.2in]{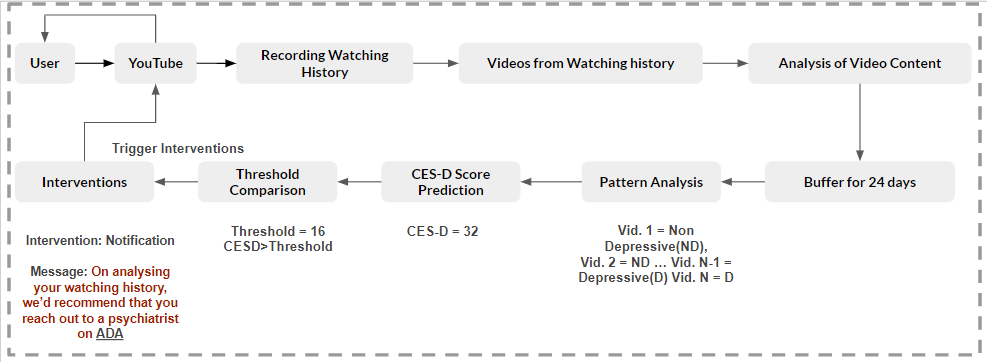}
\caption{Pipeline of the solution proposed- Detecting depression based on YouTube watching activity}
\label{fig:flow}
\end{center}
\end{figure*}

In this paper, we focus on the second layer, i.e. analyzing the video content based on the
textual features extracted from the transcripts to construct classifiers that can identify if a video can be termed as depressive or not. If, on the analysis, it is found that content of most of the videos been
watched by him/her can be classified as depressive (Pattern 3), it gives us a hint about a decline in his mental state and help us make him self-aware or direct him to the support portals if needed.

Our major contributions to this paper are:
\begin{itemize}
\item We construct a classifier that can help understand the content of a video by classifying it as Depressive/Non-Depressive.
\item To provide a real-life validation of the classification results in above step, we propose a methodology to evaluate the comments posted for a video and determine a potential score that would have been obtained on a CES-D scale and use it as a real life proxy to judge the accuracy of the classification.
\end{itemize}

\section{Data and Classification}
\subsection{Data}
The data for the analysis of the video content was gathered by extracting videos using various keywords and extracting the transcript out of it. There was a total of around 3000transcripts, totalling 1,409,719 words. The videos were divided into two categories: depressing and not-depressing. Of the data collected, 1,427( 48\%) are considered depressing and these transcripts account for nearly 754,883 words. The 1,573 (52\%) non-depressed transcripts are responsible for the other 832,117 words. To gather our data, we considered a video depressing if it was a search result when depressing keywords were used. While looking for depressing videos, we aimed mostly on videos which had tags as “self-harm”, “suicidal”, “triggering” etc. For the list of keywords, see Appendix D. Note that the keywords mentioned in the appendix are manually validated to contain depressive videos. Similarly, we collected videos around the categories like educational, funny etc. for non-depressing label.

\subsection{Classification and Results}
With respect to classification, we experimented with three models.
\begin{itemize}
\item \textbf{Naive Bayes with n-grams}: We built a Multinomial Naive Bayes (NB) Classifier using features extracted from Empath model [5]. Empath is a living lexicon mined from modern text on the web for analyzing text across lexical categories (similar to LIWC [16]). These features were the normalized score for a set of lexical categories. Apart from the pre-defined features like positive and negative emotions, the Empath model also gives the flexibility to create customized features. Therefore, categories that included seed-terms pertaining to symptoms diagnosed by the CES-D questionnaire [15] were also created and included in the set of categories. A list of seed-terms corresponding to questions is mentioned in Appendix B.

\begin{table}[h]
\begin{minipage}{.5\linewidth}
\caption{Hyperparameter details for Model 3}
\label{comparison}
\vskip 0.15in
\begin{center}
\begin{small}
\begin{sc}
\begin{tabular}{lr}
\multicolumn{1}{l}{\bf Hyperparameter}  &\multicolumn{1}{r}{\bf Value(\%) (s)}
\\ \hline \\
N-Gram range        &2-3 \\
Maximum Features  &5000 \\
\end{tabular}
\end{sc}
\end{small}
\end{center}
\vskip -0.1in
\end{minipage}\hfill
\begin{minipage}{.5\linewidth}

\caption{Model comparison}
\label{comparison}
\vskip 0.15in
\begin{center}
\begin{small}
\begin{sc}
\begin{tabular}{lcr}
\multicolumn{1}{l}{\bf Model}  &\multicolumn{1}{c}{\bf Accuracy (\%)} &\multicolumn{1}{r}{\bf Time (s)}
\\ \hline \\
Empath + NB         &52   &5 \\
TF-IDF + Empath + NB  &81.2  &8 \\
LSTM         &83.4  &345  \\
\end{tabular}
\end{sc}
\end{small}
\end{center}
\vskip -0.1in

\end{minipage}

\end{table}

\item \textbf{LSTM}: While the Naive Bayes Model provided a baseline, we used LSTM to get better results. The full structure of our LSTM model is shown in Fig.2. For each comment that's pre-processed, we make it pass through an Embedding layer seeded with pre-trained GloVe 300D word embedding weights. We feed this into a LSTM network with 196 units and tanh activation, applying a recurrent dropout of 0.2. The outputs of the LSTM are then fed through two dense layers with a dropout layer of 0.5 in between to output the video classification. We use binary cross-entropy loss and Adam optimization.

\begin{figure}[h!]
\begin{center}
\includegraphics[width = 2.5in, height = 0.7in]{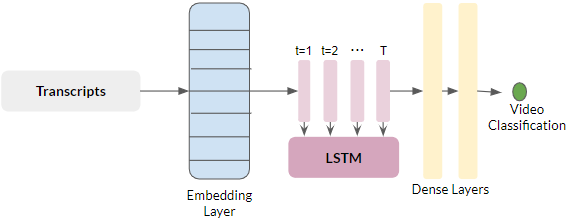}
\label{fig:LSTM} 
\caption{LSTM Model Structure}
\end{center}
\end{figure}

\item \textbf{Naive Bayes with n-grams}: We tried to modify the baseline model by capturing textual features by extracting TF-IDF weighted combinations of word ngrams from the transcripts and used it along with the features from previous model to train the Naive Bayes classifier. The hyper-parameters and their corresponding values are provided in Table 1.
\end{itemize}
A comparison of the results obtained is shown in Table 2. From the table it can be inferred that the combination of the Empath model with TF-IDF features resulted in almost the same classification accuracy as of LSTM.

\section{Evaluation}
\subsection{CES-D Scores}
For the validation and evaluation of the results produced by the classifier, we introduce a scoring method, called CES-D score, for each video to analyze how depressive the video is. The score is the density of the terms, derived from the various symptoms (Insomnia, Self-hate, Appetite, etc.) considered in the CES-D scale [15], present in a negative connotation within a given text. The Center for Epidemiologic Studies Depression Scale is a brief self-report questionnaire developed to measure depressive symptoms severity in the general population (See Appendix A). To identify the occurrence of symptoms, we use Empath Model [5] which combines modern NLP techniques to construct categories on demand using a few seed terms. For e.g., for the seed term “weight-loss”, the generated set of words were “weight loss”, “anemia”, ”stress”, “malnutrition” etc. The various seed-terms used for generation of categories and examples of the seed-terms identified from the comments' text are mentioned in Appendix B.

\subsubsection{Calculation of CES-D Score}

\begin{algorithm}[h]
\caption{CES-D Score}
\begin{algorithmic} 
\REQUIRE data $text$
\STATE set ( $categoryTerms$)
\STATE $termFreq \leftarrow 0$
\WHILE{$term$ in $categoryTerms$}
\IF{$term$ in $text$}
\STATE $termFreq \leftarrow termFreq + 1$
\ENDIF
\ENDWHILE
\STATE $termFreq \leftarrow \frac{termFreq}{len(text)}$
\STATE $neg \leftarrow Empath.analyze(text, negative)$
\STATE $pos \leftarrow Empath.analyze(text, positive)$
\STATE $connotation \leftarrow (neg-pos)$
\STATE $CESD \leftarrow termFreq \times connotation$\\
\STATE return $CESD$
\end{algorithmic}
\end{algorithm}

Algorithm 1 presents how the score is calculated for each comment in the video. The categories generated by each seed-term is merged in a set ({\it categoryTerms}). The algorithm iterates through each word of the set and calculates the {\it termfrequency} for it and calculates an aggregate by adding the term-frequency for each word. Since the words in the set can be used in both positive and negative connotations, the comment is then sent to the Empath model for analysis of a score for positive and negative emotion. The aggregate generated in the previous step is then multiplied by the difference of positive score form the negative. This is to ensure that only the negative emotions add up to the score. If for any comment, the positive score exceeds the negative score, the CES-D score is calculated as 0.\\
For the calculation of the CES-D score for the video as a whole, in order to determine the intensity of depression in the comments, the aggregate calculated in the first step is added up for all comments and a normalized score is generated by dividing the sum by the total number of comments.

\subsection{Analyzing Comments}

\begin{figure*}[hbt!]
\begin{minipage}{.5\textwidth}
\begin{center}
\includegraphics[width = \textwidth, height = 1.5in]{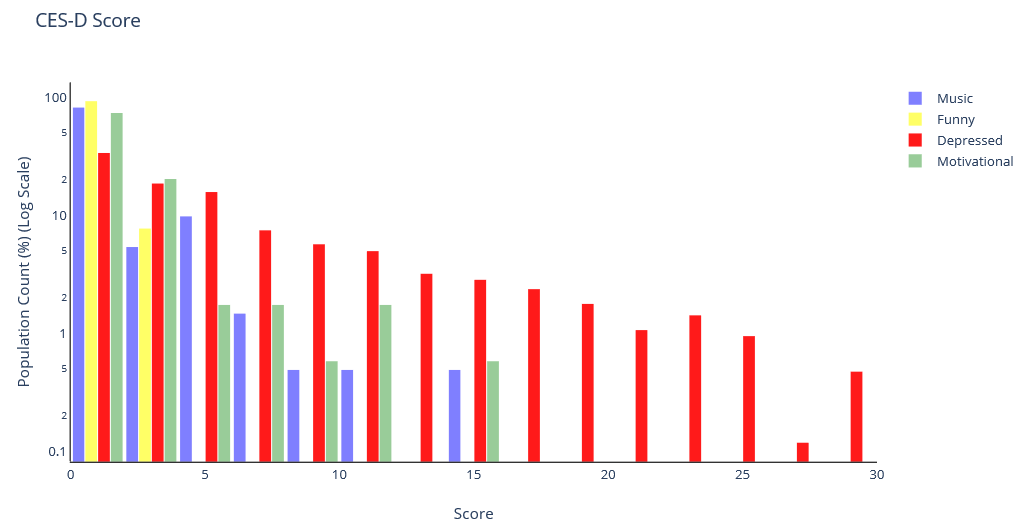}
\caption{Comparison of CES-D scores in various categories of videos}
\label{fig:CES-D}
\end{center}
\end{minipage}
\begin{minipage}{.5\textwidth}
\begin{center}
\includegraphics[width = 2.2in, height = 1.5in]{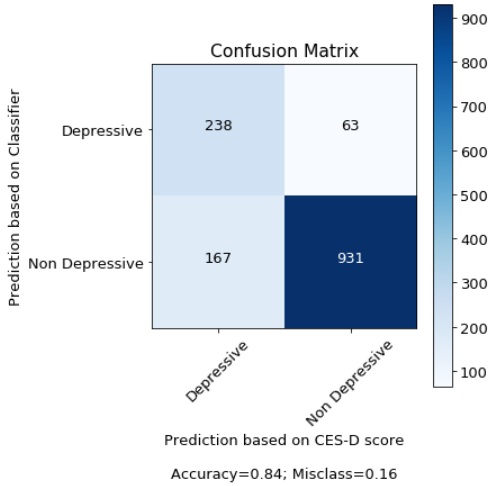}
\label{fig:matrix} 
\caption{Confusion Matrix for evaluation of classifications}
\end{center}
\end{minipage}
\end{figure*}

The comments posted on a video is a great way of understanding how the video made the viewers feel. 
They give us a fair idea of the video’s type and content. So, we examined responses to a video by analyzing the textual features of the comments considering the idea that the words a person uses reveal information about their psychological state [6]. We extracted the comments posted on videos that fall under one of the following categories:\\
\\
a) Music b) Depression c) Funny d) Self-Help/Motivational\\
\\
It was determined that a video belonged to a category if it was in the YouTube search result when the category names were used as keywords. For each keyword (category), we
extracted around 200 videos and analyzed the comments in a cumulative manner by calculating their CES-D score. On the comparison of various categories mentioned above, the difference between the average CES-D scores of depressive vs non-depressive videos was quite evident in Figure 2. In the plot, the percent of non-depressed comments (CESD = 0) for non-depressive categories (music, funny) was much higher than that in depressive category. Also, the range of the CES-D score is the largest in depressive videos which proves that depressed are induced and attracted by it. As per different theories, it has been proven that on being sad/depressed, humans tend to be inclined towards groups where they can find similarities to their mental states [8]. It can thus be deduced that watching videos of high CES-D score (depressive videos) on a frequent basis is likely to be an indicator of an unhealthy mental state.

\subsection{Evaluation Results}

To evaluate the classification results produced by our classifier in a real-life setting, we tested the results in a less constructed domain. The model was fed the transcripts of
1500 random videos and the classifications were compared to the CES-D score obtained by the comments. The method mentioned in section 3.1 was used to calculate the CES-D score of the video. A CES-D score greater than 20 was considered depressive. We came to the conclusion of setting the threshold as 20 after taking the average of CES-D scores for the depressive videos collected for the training purpose. The accuracy of the classifier based on this methodology came out to be 84\% as can be evaluated from the confusion matrix shown in Figure 4.

\section{Conclusion and Future Works}
This work demonstrates a method to analyze videos for depressive content. It also shows that individuals with a depressed state of mind tend to ruminate and watch depressive videos in order to gain validation or find company. This behavioral pattern can be useful in determining the mental state of a viewer. Since an individual should show persistent symptoms for at least two weeks to be called depressed [13], our tool will extract data over a window size of 24 days with an overlap of 10 days. The kind of videos he has watched will be analyzed for depressive content and patterns. This can be used to monitor recommendations and restrict the recommendations of triggering videos based on the watching activity. In case the patterns reveal the signs on unhealthy mental state, our tool aims to provide early diagnosis by warning the user about the change points in his mood and also direct them to the right portals for help.

The results obtained in this paper are using textual features, therefore our future work will focus on the usage of audio/visual features, and analyzing the affective content using continuous dimensions like arousal/valence, we can infer more and get better results for video content analysis. 

\section{Acknowledgements}
We would like to express our gratitude to Kris Sankaran (MILA) for his continuous support.

\section*{References}

\medskip

\small

\bibliography{neurips_2019}
\bibliographystyle{neurips_2019}

[1] Almeida, H., Briand, A., and Meurs, M.-J. Detecting early risk of depression from social media user-generated content. InCLEF, 2017.

[2] Cohn, J. F., Kruez, T. S., Matthews, I., Yang, Y., Nguyen,M. H., Padilla, M. T., Zhou, F., and De la Torre, F. Detecting depression from facial actions and vocal prosody. In {\it 2009 3rd International Conference on Affective Computing and Intelligent Interaction and Workshops}, pp. 1–7.IEEE, 2009.

[3] De Choudhury, M., Counts, S., and Horvitz, E. Predicting postpartum changes in emotion and behavior via social media. In {\it Proceedings of the SIGCHI conference on human factors in computing systems}, pp. 3267–3276. ACM,2013a

[4] De Choudhury, M., Counts, S., and Horvitz, E. Social media as a measurement tool of depression in populations.  In {\it Proceedings of the 5th Annual ACM Web Science Conference}, pp. 47–56. ACM, 2013.

[5] Fast,  E.,  Chen,  B.,  and  Bernstein,  M.  S.   Empath:  Understanding  topic  signals  in  large-scale  text.   In {\it Proceedings of the 2016 CHI Conference on Human Factors in Computing Systems},  CHI ’16,  pp. 4647–4657, New York, NY, USA, 2016. ACM.  ISBN 978-1-4503-3362-7. {\bf URL} http://doi.acm.org/10.1145/2858036.2858535.

[6] Fine, J. {\it Language in psychiatry: A handbook of clinical practice}. Equinox London, 2006.

[7] Giannakakis, G., Pediaditis, M., Manousos, D., Kazantzaki,E.,  Chiarugi,  F.,  Simos,  P.,  Marias,  K.,  and  Tsik-nakis,   M.Stress  and  anxiety  detection  using  facial  cues  from  videos. {\it Biomedical Signal Processing and Control},  31:89  –  101,  2017.ISSN  1746-8094. {\bf URL} http://www.sciencedirect.com/science/article/pii/S1746809416300805.

[8] Gray, H. M., Ishii, K., and Ambady, N. Misery loves company: When sadness increases the desire for social connectedness. {\it Personality and Social Psychology Bulletin}, 37(11):1438–1448, 2011.

[9] Guntuku, S. C., Yaden, D. B., Kern, M. L., Ungar, L. H.,and Eichstaedt, J. C.  Detecting depression and mental illness on social media:  an integrative review.{\it Current Opinion in Behavioral Sciences}, 18:43–49, 2017.

[10] Jashinsky, J., Burton, S. H., Hanson, C. L., West, J., Giraud-Carrier,  C.,  Barnes,  M.  D.,  and  Argyle,  T. Tracking suicide risk factors through twitter in the us. {\it Crisis}, 2014.

[11] Levinson, W., Gorawara-Bhat, R., and Lamb, J. A study of patient clues and physician responses in primary care and surgical settings. {\it Jama}, 284(8):1021–1027, 2000.

[12] Moreno, M. A., Jelenchick, L. A., Egan, K. G., Cox, E.,Young, H., Gannon, K. E., and Becker, T. Feeling bad on facebook: depression disclosures by college students on a social networking site. {\it Depression and anxiety}, 28(6):447–455, 2011.

[13] NIMH, D. Overview. {\bf URL} https://www.nimh.nih.gov/health/topics/depression/index.shtml.

[14] Pampouchidou, A., Simantiraki, O., Fazlollahi, A., Pedi-aditis, M., Manousos, D., Roniotis, A., Giannakakis, G.,Meriaudeau, F., Simos, P., Marias, K., et al. Depression assessment by fusing high and low level features from audio, video, and text. In {\it Proceedings of the 6th International Workshop on Audio/Visual Emotion Challenge}, pp.27–34. ACM, 2016.

[15] Radloff, L. S.  The ces-d scale:  A self-report depression scale for research in the general population. {\it Applied Psychological Measurement}, 1(3):385–401, 1977.

[16] Tausczik, Y. R. and Pennebaker, J. W.  The psychological meaning of words: Liwc and computerized text analysis methods. {\it Journal of language and social psychology}, 29(1):24–54, 2010.

[17] Yates, A., Cohan, A., and Goharian, N. Depression and self-harm risk assessment in online forums. {\it arXiv preprint arXiv:1709.01848}, 2017.

\end{document}